# Fingernail-Based Tangential Force Simulation for Enhanced Dexterous Manipulation in Virtual Reality

Yunxiu Xu[1], and Shoichi Hasegawa[1]

[1] *Tokyo Institute of Technology, Tokyo, Japan*

(Email: yunxiu@haselab.net)

**Abstract ---** This study introduces a novel haptic device for enhancing dexterous manipulation in virtual reality. By stimulating mechanoreceptors on both sides of the fingernail, our lightweight system simulates tangential force sensations. We employ mechanical stimulation for more natural tactile feedback. A preliminary "balancing grasp challenge" experiment shows that users make more frequent micro-adjustments with our device, indicating improved precision. This research aims to advance haptic feedback in VR, potentially leading to more immersive and realistic virtual interactions.

## 1 INTRODUCTION

Achieving an immersive Virtual Reality (VR) experience still faces a critical challenge: how to realize realistic dexterous grasping in virtual environments. Dexterous manipulation is a fundamental capability for human interaction with the physical world, and the lack of this ability in VR significantly limits users' sense of immersion and operational precision. To address this deficiency, hand-based haptic feedback devices have emerged as a crucial research topic. Nevertheless, existing haptic devices are often bulky and heavy, severely impeding users' freedom of movement in virtual environments. There is an urgent need to develop lightweight, miniaturized haptic devices for hand interaction. More importantly, these devices must not only be compact and lightweight but also capable of providing tactile feedback to users' fingertips that perceive real-world sensations, enabling precise manipulation and natural interaction in virtual environments. Ando et al. [5] achieved this by relocating the actuator to the lateral side of the fingernail. However, our aim is to present a more comprehensive range of tactile information without placing large, rigid actuators on the finger pad, including pressure, tangential, and physics-based vibration feedback.

In the process of dexterous grasping, tangential forces play a crucial role. When we grasp objects, due to the deformation of the finger pad's skin, the fingertips not only sense vertical pressure but also perceive tangential forces on the object's surface, which may be essential for precise control and perception of the grasping state. However, existing haptic devices [3] [4] that attempt to reproduce tangential force feedback often struggle to maintain the freedom of the finger pad. This not only affects user comfort but may also interfere with the tactile feedback of the real world.

Our focus is on stimulating the receptors located on both sides of the fingernail. Previous research has established that the SA-II mechanoreceptors in the fingernail walls reliably respond to force, providing direction information about fingertip forces to the central nervous system [7]. Nakayama et al. [1] implemented this concept using electrical stimulation methods. However, compared to electrical stimulation, mechanical stimulation produces more natural sensations, also with smaller individual variations. Our aim is to investigate whether applying mechanical stimulation to both sides of the fingernail can effectively simulate tangential force sensations and potentially enhance dexterous manipulation. In contrast to the approach of Minamizawa et al. [2], this study is grounded in physiological findings related to mechanoreceptors around the fingernail, potentially aligning more closely with natural tactile perception mechanisms.

The proposed device combines mechanical stimulation of the fingernail's lateral sides with a lightweight design. We proposed using mechanical stimulation to activate SA-II receptors in collagen fibers connected to the nail wall[7]. Unlike existing haptic devices, it provides tangential force feedback without restricting finger pad movement using a single motor and can be integrated with other haptic displays, enabling more precise, dexterous manipulation in virtual environments.

## 2 SYSTEM OVERVIEW

We selected the PIC32 microcontroller, which can output multiple channels of PWM signals. The microcontroller receives haptic information from the PC. We paired this with a motor driver featuring the DRV8434e current controller, allowing us to regulate each motor's output current. To display tangential force, we utilized a coreless motor placed on the dorsal side of the fingernail, which drives a 'horn-shaped' structure, as shown in Fig. 1. This structure features two contact points designed to stimulate the mechanoreceptors located in the skin folds adjacent to the fingernail. The force magnitude is controlled by adjusting the motor torque. Additionally, since the structure produces a sharp collision when it impacts the fingernail, we used shock-absorbing material to reduce vibrations. Compared to the mechanoreceptors on the palmar skin in direct contact with the object's surface, the force signals provided by SA-II receptors in the nail may be less influenced by the geometric and textural features of the contact surface. Therefore, to deliver fingertip pressure, we employed the previous structure [6] using fine wires to achieve both pressure and vibration. This also demonstrates that our tangential force display structure is easily integrated with other tactile displays.

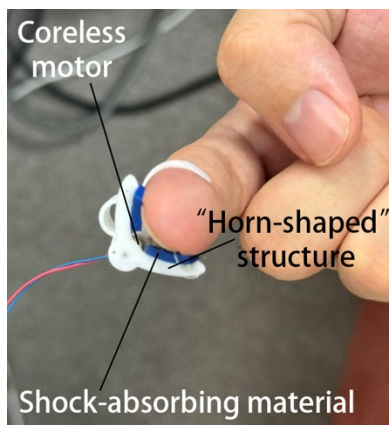

Fig. 1 Hardware overview.

## 3 PRELIMINARY EXPERIMENT

We asked 3 participants to conduct an experiment to evaluate the effectiveness of our implemented tangential force in physics-based hand manipulation. The experiment utilized a feedback pattern where compression of the left side of the fingernail occurred when the finger pad slid from right to left on a flat surface, and compression of the right side occurred when the fingertip slid from left to right. The task can be described as a "balancing grasp challenge," where participants needed to use their thumb and index finger to pick up a small rectangular block placed on a non-horizontal surface and maintain the block in a horizontal position as much as possible. This meant keeping the long edge of the block parallel to the ground after grasping, avoiding noticeable tilting. During the task, the participants wore the proposed device on their thumb and index finger, and the system recorded the block's tilt angle in real-time. The experiment was conducted under two conditions: with tangential force haptics enabled and disabled.

The results indicated that with tangential force feedback, all three participants exhibited multiple small, repeated adjustment actions while performing the task, as shown in Fig. 2. These actions manifested as periodic, smooth micro-adjustments of the fingers while grasping and maintaining the object. This phenomenon was not observed when haptics was disabled, suggesting that participants may have been using tactile information for more precise control, rather than relying solely on visual cues. However, the current experimental results might reflect participants' adaptation to the tactile feedback rather than genuine fine motor control. Further research is needed to confirm whether the haptic feedback accurately simulates human tactile perception.

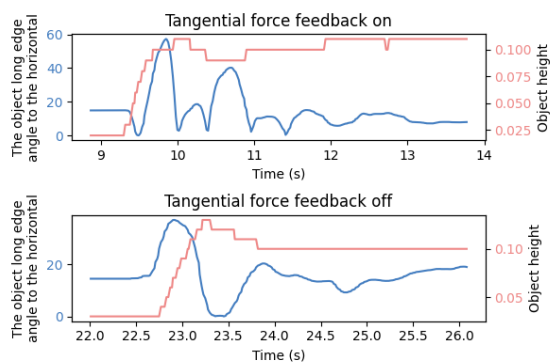

Fig. 2 Result of a typical user: After grasping an virtual object on a slope, the participant was asked adjusted it to horizontal. With haptics enabled, more frequent fine adjustments were observed.

## 4 CONCLUSION

This study introduces a novel haptic device for simulating tangential forces on fingernails, potentially enhancing dexterous manipulation in virtual reality. Our preliminary experiment suggests that this approach enables more precise object control through micro-adjustments. In future work, we aim to investigate whether this device can enhance the efficiency of physics-based manipulation, and find solution for display tangential force in multiple directions.

## 5 Future Works

We plan to increase the number of participants and measure whether directional perception aligns with expectations. Subsequently, we will use physics-based tasks, such as object weight recognition and surface friction coefficient identification, to quantitatively assess whether directional haptic feedback enhances user recognition capabilities. The proposed device still has limited degrees of freedom and cannot present tactile feedback in all planar directions. Future research will focus on using linear actuators to push and pull the fingernail along the proximal-distal axis.

## 6 Acknowledgements

This work was supported by JST SPRING, Japan Grant Number JPMJSP2106, and JSPS KAKENHI Grant Number 23H03432.